\documentclass[12pt]{iopart}
\usepackage{iopams}
\usepackage{graphicx}
\usepackage{amsfonts,bm,amssymb}
\usepackage{color}

\begin{document}

\title{Exactness of Belief Propagation\\  for Some Graphical
Models with Loops}

\author{Michael Chertkov}
\address{\mailto{chertkov@lanl.gov}\\Theoretical Division and Center for Nonlinear
Studies, Los Alamos National Laboratory, Los Alamos, NM 87545}

\begin{abstract}
It is well known that an arbitrary graphical model of statistical inference defined
on a tree, i.e. on a graph without loops, is solved exactly and efficiently by an
iterative Belief Propagation (BP) algorithm convergent to unique minimum of the
so-called Bethe free energy functional. For a general graphical model on a loopy
graph the functional may show multiple minima, the iterative BP algorithm may
converge to one of the minima or may not converge at all, and the global minimum of
the Bethe free energy functional is not guaranteed to correspond to the optimal
Maximum-Likelihood (ML) solution in the zero-temperature limit. However,  there are
exceptions to this general rule, discussed in \cite{05KW} and \cite{08BSS} in two
different contexts, where zero-temperature version of the BP algorithm finds ML
solution for special models on graphs with loops. These two models share a key
feature: their ML solutions can be found by an efficient Linear Programming (LP)
algorithm with a Totally-Uni-Modular (TUM) matrix of constraints. Generalizing the
two models we consider a class of graphical models reducible in the zero temperature
limit to LP with TUM constraints.  Assuming that a gedanken algorithm, g-BP, finding
the global minimum of the Bethe free energy is available we show that in the limit
of zero temperature g-BP outputs the ML solution. Our consideration is based on
equivalence established between gapless Linear Programming (LP) relaxation of the
graphical model in the $T\to 0$ limit and respective LP version of the Bethe-Free
energy minimization.
\end{abstract}


 \submitto{Journal of Statistical Mechanics}


\maketitle

\section{Introduction}

Belief Propagation (BP) is an algorithm finding ML solution or marginal probabilities on a graph
without loops, a tree. The algorithm was introduced in  \cite{63Gal} as
an efficient heuristic for decoding of sparse (so called graphical) codes and it
was independently considered in the context of graphical models of artificial
intelligence \cite{88Pea}. Originally the algorithm was primarily thought of as an
iterative procedure. \cite{01YFW,05YFW}, inspired by earlier works of \cite{35Bet}
and  \cite{36Pei} in statistical physics, suggested to use a more fundamental notion
of the Bethe free energy. Extrema of the Bethe free energy represent fixed points of
the iterative BP algorithm on graphs with cycles. Equations describing the
stationary points of the Bethe free energy and the fixed points of the iterative BP
are called Belief Propagation, or Bethe-Peierls, BP equations.

The significance of BP, understood as an algorithm looking for a
minimum of the Bethe free energy, was further elucidated within the
framework coined loop calculus \cite{06CCa,06CCb}. It was shown that algorithm finding an extremum of the
Bethe Free Energy is not just an approximation/heursistics in the loopy case, but it
allows explicit reconstruction of the exact inference in terms of a
series,  where each term corresponds to a loop on the graph.

If the graphical model is dense there are many loops, and thus many
contributions to the loop series. However, not all loops are equal.
Thus, considering models characterized by the same graph but
different factor functions or local weights (exact definitions will
follow) one expects strong sensitivity of an individual loop
contribution (and its significance within the loop series) on the factor
functions. In this context it is of interest to study the following question:
are there graphical models, defined on an arbitrary graph
but with specially tuned factor functions, such that BP provides
exact inference?

Positive answer to this question was given, independently and for two different
models, by  \cite{05KW} and \cite{08BSS}. It was shown in \cite{05KW} that for a
graphical model defined on an arbitrary graph in terms of binary variables with
pairwise sub-modular interaction a properly defined version of BP (linear
programming relaxation underlying the tree-reweighted method of \cite{03WJW})
yields a globally optimal Maximum Likelihood solution. This model is equivalent to
the Ferromagnetic Random Field Ising (FRFI) model popular in statistical physics of
disordered systems, see e.g. \cite{02HR}. Maximum weight matching problem on a bipartite graph
was analyzed in \cite{08BSS} and later in \cite{07SMW}, where it was shown that, in spite of the fact that the
underlying graph has many short cycles, an algorithm of BP type does converge to
correct ML assignment. This consideration was also extended to the problem of weighted b-matchings on an arbitrary graph (which is
yet another problem solvable exactly by BP) in \cite{07BBCZ,07HJ}.
Closely related general results, discussing convexified version of Bethe Free energy  and
an iterative convex-BP scheme converging to respective LP, were reported in \cite{07WYM,07GJ}.

\begin{figure}
\begin{center}
\includegraphics[width=12cm]{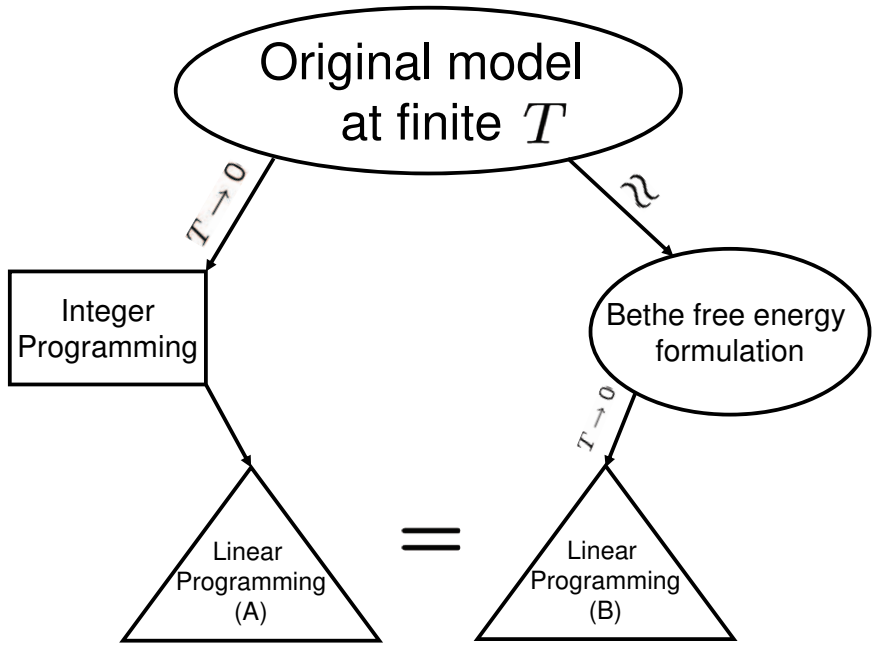}
\end{center}
\caption{Scheme illustrating the sequence of transformations and
relations discussed in the paper. } \label{fig:scheme}
\end{figure}

In this paper we use the Bethe free energy approach of \cite{05YFW} to suggest a
complementary and unifying explanation to these remarkable, and somehow surprising, results of
\cite{05KW,08BSS}. In two subsequent Sections we consider two models, FRFI discussed
in \cite{05KW} and a binary model with Totally Uni-Modular (TUM) constraints
generalizing the weight matching problem considered in \cite{08BSS}. Statistical
weights are defined for both models in terms of a characteristic temperature, $T$.
Our strategy in dealing with both models is illustrated in Fig.~\ref{fig:scheme}. It
consists of the following three steps.
\begin{itemize}
\item
Starting from the original setting we first go anti-clockwise,
getting an Integer Programming (IP) formulation for the ML, $T\to0$,
version of the problem. The most important feature of the two models
is that the LP relaxation of the respective IP, shown as LP-A in the
Figure, is tight/exact. In both cases this reduction from IP to LP
is exact due to the Total-Uni-Modularity (TUM) feature of the
underlying matrix of constraints.

\item Then we return to the
original setting and start moving clockwise (see the Figure), first to the Bethe
free energy formulation of the problem. We call the gedanken algorithm,  finding global
minima of the Bethe free energy, g-BP \footnote{The Bethe free energy is non-convex,
therefore funding the global minimum at a finite temperature is not necessarily
straightforward. Acknowledging importance of the problem,  we will not discuss in this manuscript plausibility and
details of respective iterative algorithm convergent to the global minimum of the Bethe free energy.
We refer interested reader to comprehensive discussion of such iterative
schemes in the general context in \cite{07WYM,07GJ} and for FRFI model and maximum weight matching model in \cite{05KW} and in \cite{08BSS,07SMW} respectively.} In the $T\to 0$ limit the
Bethe free energy turns to respective self-energy (the entropy term multiplied by
temperature is irrelevant) which is a linear functional of beliefs. Thus one gets to
an LP problem here as well, the one shown as LP-B in the Figure. This transformation
from g-BP to LP-B is analogous to similar relation between g-BP and LP-decoding
introduced in the coding theory in \cite{05FWK,03WJ}.

\item Finally, we show that LP-A and LP-B are identical, thus
demonstrating that g-BP in the $T\to 0$ limit outputs the ML solution.
\end{itemize}
Note, that convexity of the Bethe free energy at finite temperature, playing the key role in analysis of \cite{05KW,07WYM,07GJ}, is not a required
part our consideration. Moreover, the Bethe free energy of the binary model with TUM constraints is generally not convex.

\section{Ferromagnetic Random Field Ising model}

Consider an  undirected graph ${\cal G}$,  consisting of $n$
vertexes, ${\cal V}=\{1,\cdots,n\}$ and weighted edges, ${\cal E}$,
with the weight matrix $(J_{ij}|i,j=1,\cdots,n)$ such that whenever
the two vertexes are connected by an edge, i.e. $i\in j$ or $j\in
i$, $J_{ij}> 0$, and $J_{ij}=0$, otherwise. It is also useful to
introduce the directed version of the graph, ${\cal G}_d$, where any
undirected edge of ${\cal G}$ is replaced by two directed edges of
${\cal G}_d$, with the weights $J_{i\to j}=J_{j\to i}=J_{ij}/2$
respectively. The Ferromagnetic Random Field Ising (FRFI) model is
defined by the following statistical weight associated with any
configuration of ${\bm \sigma}=(\sigma_i=\pm 1|i=1,\cdots,n)$ on
${\cal V}$:
\begin{eqnarray}
 && p({\bm\sigma})=Z^{-1}\exp\left(
 \frac{1}{2T}\sum_{(i,j)\in{\cal G}}
 J_{ij}\sigma_i\sigma_j+\frac{1}{T}\sum_i
 h_i\sigma_i\right)\nonumber\\ && =Z^{-1}\exp\left(
 \frac{1}{2T}\sum_{(i\to j)\in{\cal G}_d}
 J_{i\to j}\sigma_i\sigma_j+\frac{1}{T}\sum_i h_i\sigma_i\right), \label{psigma}
\end{eqnarray}
where $h_i$ can be positive or negative; $T$ is the temperature; $Z$
is the partition function, enforcing the probability normalization
condition, $\sum_{\bm\sigma}p({\bm\sigma})=1$; and $(i,j)$/$(i\to
j)$ marks an undirected/directed edge of ${\cal G}$/${\cal G}_d$
connecting the two neighbors $i$ and $j$.

\subsection{From FRFI to the Min-Cut Problem}

Maximum Likelihood (ground state) solution of Eq.~(\ref{psigma})
turns into the problem of quadratic integer programming
\begin{eqnarray}
\left.\min_{\bm\sigma}\left(-\frac{1}{2}\sum_{(i\to j)\in{\cal
G}_d}J_{i\to j}\sigma_i\sigma_j-\sum_{i\in{\cal G}_d}
h_i\sigma_i\right)\right|_{\forall\ i\in{\cal G}_d:\ \
\sigma_i=\pm1}. \label{Quadr1}
\end{eqnarray}
It is well known that any sub-modular energy function (and the
quadratic function in Eq.~(\ref{Quadr1}) is of this type) can be
minimized in polynomial time by reducing the task to the maximum
flow/min-cut problem \cite{02BH,02HR}. In this Subsection we will
reproduce these known results.

To unify linear and quadratic terms in Eq.~(\ref{Quadr1}), one
constructs a new graph, ${\cal G}'_d$, adding two new nodes to ${\cal G}_d$:
source (s) and destination (d), with $\sigma_s=+1$ and $\sigma_d=-1$
respectively. The (s)-node is linked to all the nodes of ${\cal G}_d$ with
$h_i>0$, while any node of ${\cal G}_d$ with $h_i<0$ is linked to (d).
Weights of the newly introduced directed edges of ${\cal G}'_d$ are
\begin{eqnarray}
 J_{s\to i}=
 2h_i, \mbox{ if } h_i>0;
 \quad
 J_{i\to d}= 2|h_i|, \mbox{ if }
 h_i<0.
 \label{sd}
\end{eqnarray}
This results in the following version of Eq.~(\ref{Quadr1})
\begin{eqnarray}
\left.\min_{\bm\sigma}\left(-\frac{1}{2}\sum_{(i\to j)\in{\cal
G}'_d}J_{i\to j}\sigma_i\sigma_j\right)\right|_{\forall\ i\in{\cal
G}_d:\ \ \sigma_i=\pm1; \sigma_s=+1;\sigma_d=-1}. \label{Quadr2}
\end{eqnarray}

Reduction from quadratic integer programming (\ref{Quadr2}) to an
integer linear programming is the next step. This is achieved via
transformation to the edge variables,
\begin{eqnarray}
 \forall (i\to j)\in{\cal G}_d:\quad \eta_{i\to j}=\left\{\begin{array}{cc}
 1,& \sigma_i=1,\sigma_j=-1\\
 0,& \mbox{otherwise}.
\end{array}\right. \label{pij}
\end{eqnarray}
The relations can also be restated
\begin{eqnarray}
&& \forall (i\to j),(j\to i)\in {\cal G}_d:\quad
\sigma_i\sigma_j+\sigma_j\sigma_i=2-4(\eta_{i\to j}+\eta_{j\to i})
\label{sigma-eta}\\
 && \forall (d\to i),(j\to t):\quad
 \sigma_i\sigma_d=1-2\eta_{d\to i},\quad \sigma_s\sigma_j=1-2\eta_{j\to
 s},\label{sigma-eta_sd}
\end{eqnarray}
Therefore,  taking into account that ,  $J_{i\to j}=J_{j\to i}$ for
any $(i\to j),(j\to i)\in{\cal G}_d$, substituting
Eqs.~(\ref{pij},\ref{sigma-eta},\ref{sigma-eta_sd}) into
Eq.~(\ref{Quadr2}) and changing variables from $\sigma_i=\pm 1$ to
$p_i=(1-\sigma_i)/2=0,1$ one arrives at
\begin{equation}
 \hspace{-2cm}-\frac{1}{2}\sum\limits_{(i\to j)\in{\cal G}'_d}J_{i\to j}+
 \left.\min\limits_{\{\eta,p\}}\sum_{(i\to j)\in{\cal
 G}'_d}J_{i\to j} \eta_{i\to j}\right|_{
 \begin{array}{c}
 \forall i\in{\cal G}'_d, p_i=0,1;\\
 \forall (i\to j)\in{\cal G}'_d:\
 \ p_i-p_j+\eta_{i\to j}=0,1;\\
 p_s=0,\quad p_d=1
 \end{array} }.
 \label{min_cutIP_d}
\end{equation}
This expression is nothing but the integer programming formulation
of the famous min-cut problem, calculating the minimum weight cut
splitting all the nodes of the directed graph into two parts such
that the group including the source node has all variables in the
$0$ state while the other group, including the destination node, has
all variables in the $1$ state.

Any $\{\eta,p\}$ configuration which satisfies conditions in
Eq.~(\ref{min_cutIP_d}) requires that either $\eta_{i\to j}=0$ and $\eta_{j\to i}=1$
or $\eta_{i\to j}=1$ and $\eta_{j\to i}=0$ for any pair of directed edges $(i\to
j),(j\to i)\in {\cal G}_d$. This suggests that Eq.~(\ref{min_cutIP_d}) can  be restated in
terms of the undirected graph $G'$, equivalent to the original ${\cal G}$ supplemented by
the source and destination vertexes and edges with the following positive weights
\begin{eqnarray}
 J_{si}=
 2h_i, \mbox{ if } h_i>0;
 \quad
 J_{id}= 2|h_i|, \mbox{ if }
 h_i<0.
 \label{sd_u}
\end{eqnarray}
One derives the following undirected version of
Eq.~(\ref{min_cutIP_d})
\begin{equation}
 \hspace{-2cm}-\frac{1}{2}\sum\limits_{(i,j)\in{\cal G}'}J_{ij}+
 \left.\min\limits_{\{\eta,p\}}\sum_{(i,j)\in{\cal
 G}'}J_{ij} \eta_{i\to j}\right|_{
 \begin{array}{c}
 \forall i\in{\cal G}', p_i=0,1;\\
 \forall (i,j)\in{\cal G}':\
 \ p_i-p_j+\eta_{ij}=0,1;\\
 p_s=0,\quad p_d=1
 \end{array} }.
 \label{min_cutIP}
\end{equation}

The min-cut problem (\ref{min_cutIP}) is solvable in polynomial
time. This means, in particular, that solution of the Integer
Programming Eq.~(\ref{min_cutIP}) and solution of the respective
relaxed LP-A,
\begin{eqnarray}
 -\frac{1}{2}\sum\limits_{(i,j)\in{\cal G}'}J_{ij}+
 \left.\min\limits_{\{\eta,p\}}\sum_{(i,j)\in{\cal
 G}'}J_{ij} \eta_{ij}\right|_{
 \left\{\begin{array}{c}
 \forall i\in{\cal G}', 1\geq p_i\geq 0;\\
 \forall (i,j)\in{\cal G}':\
 p_i-p_j+\eta_{ij}\geq 0;\\
 p_s=0,\quad p_d=1
 \end{array} \right.},
 \label{min_cutLP}
\end{eqnarray}
are identical. The tightness of the relaxation is, e.g., discussed
in \cite{98PS}. (See Ch. 6.1 and specifically Theorems 6.1,6.2 in
\cite{98PS}.) Also, this observation is closely related to the fact
that the matrix of constraints in the max-flow problem,  which is
dual to Eq.~(\ref{min_cutIP}), is Totally Uni-Modular (TUM), i.e.
such that any square minor of the matrix has determinant which is
$0,+1$ or $-1$. (See e.g. Ch. 13.2 of \cite{98PS} for discussion of
the TUM IP/LP problems.)

\subsection{Bethe Free Energy and Belief Propagation for FRFI}

Discussing the FRFI model defined in Eq.~(\ref{psigma}) and
following the general heuristic approach to the graphical models,
suggested in \cite{05YFW}, one introduces beliefs, i.e. estimated
probabilities, for vertexes and edges, $b_i(\sigma_i)$,
$b_{ij}(\sigma_i,\sigma_j)$, related to each other according to
\begin{eqnarray}
\forall i\ \ \&\ \ \forall j\in i:\quad
b_i(\sigma_i)=\sum_{\sigma_j}b_{ij}(\sigma_i,\sigma_j), \label{bb}
\end{eqnarray}
and also satisfying the obvious normalization condition
\begin{eqnarray}
\forall i:\quad \sum_{\sigma_i}b_i(\sigma_i)=1. \label{norm}
\end{eqnarray}

Then the Bethe free energy functional of the beliefs is defined as
\begin{eqnarray}
 {\cal F}=E-T S,\quad E=-\frac{1}{2}\sum_{(i,j)}\sum_{\sigma_i,\sigma_j}
 b_{ij}(\sigma_i,\sigma_j)J_{ij}\sigma_i\sigma_j
 -\sum_i\sum_{\sigma_i}b_i(\sigma_i)h_i\sigma_i,\label{En}\\
 S=\sum_{(i,j)}\sum_{\sigma_i,\sigma_j}b_{ij}(\sigma_i,\sigma_j)
 \ln b_{ij}(\sigma_i,\sigma_j)
 -\sum_i\sum_{\sigma_i}b_i(\sigma_i)\ln b_i(\sigma_i).
 \label{Ent}
\end{eqnarray}

Introducing Lagrangian multipliers associated with the constraints
(\ref{bb},\ref{norm}), one defines the Lagrangian functional
\begin{equation}
 \hspace{-2cm}{\cal L}={\cal F}+\sum_i\sum_{j\in i}
 \sum_{\sigma_i}\eta_{ij}(\sigma_i)\left( b_i(\sigma_i)-\sum_{\sigma_j}b_{ij}(\sigma_i,\sigma_j)\right)
 +\sum_i\lambda_i\left(\sum_{\sigma_i}b_i(\sigma_i)-1\right).
\label{Lagr}
\end{equation}
Looking for the stationary point of the Lagrangian over all the
parameters (the beliefs and the Lagrangian multipliers) will define
the Belief Propagation (BP) equations. Iterative solution of the BP
equations constitutes the celebrated BP algorithm, which is often
used as an efficient heuristic for estimating marginal
probabilities in sparse graphical models.

\subsubsection{Ground State}

In the $T\to 0$ limit the entropy terms in the expression for the
Bethe free energy in Eqs.~(\ref{En}) can be neglected and the task
of finding the absolute minimum  of the Bethe free energy functional
turns into minimization of  the self-energy, $E$ from
Eq.~(\ref{En}), under the set of constraints (\ref{bb},\ref{norm}).
Both the optimization functional and the constraints are linear in
the beliefs, therefore one gets here the following Linear
Programming optimization:
\begin{equation}
\hspace{-2.6cm}\left.\min\limits_{\{b_i;b_{ij}\}} \left(-\!\!\!\!\sum_{(i,j)\in{\cal
G}}\sum_{\sigma_i,\sigma_j}
b_{ij}(\sigma_i,\sigma_j)\frac{J_{ij}}{2}\sigma_i\sigma_j
 \!-\!\sum_{i\in{\cal G}}\sum_{\sigma_i}b_i(\sigma_i)h_i\sigma_i\right)\right|_{
 \begin{array}{c}\forall i,(i,j)\in{\cal G}:\\
b_i(\sigma_i)=\sum_{\sigma_j}b_{ij}(\sigma_i,\sigma_j)\\
\sum_{\sigma_i}b_i(\sigma_i)=1\end{array}}, \label{LP1}
\end{equation}
where it is also assumed that all the beliefs are positive and
smaller than or equal to unity (as we are looking only for
physically sensible solutions of the optimization problem).

Making the transformation from the original graph ${\cal G}$ to its
extended version, ${\cal G}'$, i.e. introducing new edges with
weights defined in Eqs.~(\ref{sd_u}), and requiring that the spin
values of the source/destination are fixed to
 $\pm 1$ respectively, i.e. $b_{s}(+)=b_d(-)=1$ and
 $b_s(-)=b_d(+)=0$, one rewrites Eq.~(\ref{LP1}) as
\begin{equation}
\hspace{-2.6cm}\left.\min\limits_{\{b_i;b_{ij}\}} \left(-\sum_{(i,j)\in{\cal
G}'}\sum_{\sigma_i,\sigma_j}
b_{ij}(\sigma_i,\sigma_j)\frac{J_{ij}}{2}\sigma_i\sigma_j\right)\right|_{
 \begin{array}{c}\forall i,(i,j)\in{\cal G}':\quad
b_i(\sigma_i)=\sum_{\sigma_j}b_{ij}(\sigma_i,\sigma_j)\\
\forall i\in{\cal G}':\quad
\sum_{\sigma_i}b_i(\sigma_i)=1\\
b_s(+)=1\ \ \& \ \ b_d(-)=1
\end{array}}. \label{LP2}
\end{equation}

The structure of the optimization functional in Eq.~(\ref{LP2})
suggests to reduce the number of variables (beliefs), thus keeping
only the edge variables
\begin{eqnarray}
 \mu_{ij}\equiv b_{ij}(+,-)+b_{ij}(-,+)=1-b_{ij}(+,+)-b_{ij}(-,-),\label{muij}
\end{eqnarray}
defined  as the probabilities to observe the edge $(i,j)$ either in
the state $(+,-)$ or in the state $(-,+)$.  Thus,  by construction,
$1\geq \mu_{ij}\geq 0$. The $\mu_{ij}$ variables defined at
different edges are related to each other through local beliefs,
$\pi_i=b_i(-)=1-b_i(+)$, which all satisfy, $0\leq \pi_i\leq 1$.
Taking all these observations into account one rewrites
Eq.~(\ref{LP2}) as
\begin{equation}
\hspace{-2.6cm} -\frac{1}{2}\sum\limits_{(i,j)\in{\cal
G}'}\!\!J_{ij}+\left.\min\limits_{\{\mu,\pi\}}\sum_{(i,j)\in{\cal G}'}J_{ij}
\mu_{ij}\right|_{
 \begin{array}{c}
 \forall i\in{\cal G}',\ \ \forall j\in i:\
 \pi_i-\pi_j+\mu_{ij}\geq 0;\ \ \ 1\geq\mu_{ij}\geq 0\\
 \forall i\in{\cal G}':\quad 1\geq \pi_i\geq 0\\
 \pi_s=0,\quad \pi_d=1
 \end{array} }.
 \label{LP3}
\end{equation}
One finds that, up to an obvious change of variables from $\mu$ to
$\eta$ and from $\pi$ to $p$, the LP-B of Eq.~(\ref{LP3}) is
identical to the LP-A (\ref{min_cutLP}). According to the Theorem
6.1 of \cite{98PS}, solutions of Eq. (\ref{LP3}), or
Eq.~(\ref{min_cutLP}), are integers, $\forall (i,j)\in\Gamma'$
$\mu_{ij},\eta_{ij}=0,1$ and $\forall i\in G'$ $p_i,\pi_i=0,1$.

Summarizing, it was just shown that as $T\to 0$ the BP solution of
the FRFI model, understood as the global minimum of the Bethe Free
energy, is also the ML solution of the model.

\section{Binary model with Totally Uni-Modular Constraints}

Consider $N$ binary variables combined in the vector
${\bm\sigma}=(\sigma_i=0,1|i=1,\cdots,N)$, and associate the
following normalized probability to any possible value of the vector
\begin{eqnarray}
p({\bm\sigma})=Z^{-1}\exp\left(-T^{-1}\sum_i
h_i\sigma_i\right)\prod\limits_\alpha\delta\left(\sum_iJ_{\alpha
i}\sigma_i,m_\alpha\right),\label{ps1}
\end{eqnarray}
where $\delta(x,y)$ is one if $x=y$ and it is zero otherwise;
$\alpha=1,\cdots,M$ ; matrix
$\hat{J}\equiv\left(J_{\alpha
i}=0,1|i=1,\cdots,N;\alpha=1,\cdots,M\right)$ is Totally Uni-Modular
(TUM),  i.e.  determinant of any square minor
of the matrix is $0,1$ or $-1$; the vector ${\bm
m}=(m_\alpha|\alpha=1,\cdots,M)$ is constructed from positive
integers, so that $\forall\alpha$: $m_\alpha\leq
q_\alpha\equiv\sum_i J_{\alpha i}$. The partition function $Z$
is introduced in Eq.~(\ref{ps1}) to guarantee normalization, $\sum_{\bm\sigma}
p({\bm\sigma})=1$.

The model Eq.~(\ref{ps1}) can be viewed as a graphical model defined on the
bipartite graph consisting of ``bits", $\{i\}$, and ``checks", $\{\alpha\}$. Also,
there may be other graphical interpretations. Thus,  for the weighted matching
problem,  e.g. studied in \cite{08BSS}, the binary variables in the formulation of
Eq.~(\ref{ps1}), are associated with edges of the complete bipartite graph. (In this
case of the weighted matching, one can show that the resulting matrix of constraints
is indeed TUM.)

\subsection{Efficient ML solution}

We, first of all, observe that the problem of finding the Maximum
Likelihood of Eq.~(\ref{ps1}) is equivalent to the following Integer
Programming (IP)
\begin{eqnarray}
 \left.\min_{\bm\sigma}\sum_i h_i\sigma_i\right|_{
 \left\{\begin{array}{c}
 \forall i:\ \ \sigma_i=0,1\\
 \forall \alpha:\ \ \sum_i J_{\alpha i}\sigma_i=m_\alpha
 \end{array}\right.}.
 \label{IP1}
\end{eqnarray}

Relaxing the IP to respective LP-A. with $\sigma_i=0,1$ changed to
$s_i=[0;1]$,
\begin{eqnarray}
 \left.\min_{\bm\sigma}\sum_i h_i s_i\right|_{
 \left\{\begin{array}{c}
 \forall i:\ \ 0\leq s_i\leq 1\\
 \forall \alpha:\ \ \sum_i J_{\alpha i} s_i=m_\alpha
 \end{array}\right.}.
 \label{LPA1}
\end{eqnarray}
one finds that the relaxation is tight. In other words,  the
solutions of the IP problem and the LP problem are exactly
equivalent. This is due to the Theorem (see e.g. Theorem 13.1 of
\cite{98PS}) stating that if $\hat{J}$ is TUM and ${\bm m}$ is
integer, then all feasible solutions of the LP problem are integer.

\subsection{Bethe Free Energy \& BP}

Here we discuss the Bethe free energy/Belief Propagation (BP)
approach to the model defined in Eq.~(\ref{ps1}). The Bethe free
energy functional is
\begin{eqnarray}
&& F=E-TS,\quad E=\sum_i h_i b_i(1),\label{FE_TUM}\\ &&
S=\sum_\alpha\sum_{{\bm\sigma}_\alpha}b_\alpha({\bm\sigma}_\alpha)\ln
b_\alpha({\bm\sigma}_\alpha)-\sum_i\left(q_\alpha-1\right)b_i(\sigma_i)\ln
b_i(\sigma_i), \nonumber
\end{eqnarray}
where a vector ${\bm\sigma}_\alpha\equiv(\sigma_i|\forall i\mbox{
s.t. } J_{\alpha i}=1;\sum_i J_{\alpha i}\sigma_i=m_\alpha)$ defines
the set of allowed configurations of variables marked by index $i$
associated and consistent with the given constrained $\alpha$. For
any given $m_\alpha$ the number of such allowed
vectors/configurations of ${\bm\sigma}_\alpha$ is
$C_{m_\alpha}^{q_\alpha}=m_\alpha
!/((m_\alpha-q_\alpha)!q_\alpha!)$. As usual,
$b_\alpha({\bm\sigma}_\alpha)$ and $b_i(\sigma_i)$ are beliefs
(estimations for the respective probabilities) associated with the
variables and the constraints. The two types of beliefs are related
to each other via the following compatibility constraints:
\begin{eqnarray}
\forall i\ \ \&\ \ \forall \alpha\mbox{ s.t. }  J_{\alpha i}=1:\ \
b_i(\sigma_i)=\sum_{{\bm\sigma}_\alpha\setminus\sigma_i}b_\alpha({\bm\sigma}_\alpha),
\label{compL}
\end{eqnarray}
and one should also impose the normalization constraint
\begin{eqnarray}
\forall\ \ i:\ \  \sum_{\sigma_i} b_i(\sigma_i)=1. \label{normL}
\end{eqnarray}

Incorporating the compatibility and normalization constraints in the
form of Lagrangian multipliers into the variational functional one
derives the Lagrangian
\begin{equation}
 \hspace{-2cm}{\cal L}=F+\sum_i \sum_{\alpha\ni i}\mu_{\alpha i}(\sigma_i)
 \left(b_i(\sigma_i)-\sum_{{\bm\sigma}_\alpha\setminus\sigma_i}b_\alpha({\bm\sigma}_\alpha)\right)+
 \sum_i\lambda_i(\sigma_i)\left(\sum_{\sigma_i} b_i(\sigma_i)-1\right).
\label{LagrL}
\end{equation}
Looking for the stationary points of the Lagrangian with respect to
all the beliefs and the Lagrangian multipliers, $\lambda$,$\mu$, one
arrives at the Belief Propagation equations for the problem.

\subsection{$T\to 0$ limit of the Bethe free energy}

In the $T\to 0$ limit the entropy term in Eq.~(\ref{FE_TUM}) can be
dropped and the problem turns into minimization of the LP type
\begin{eqnarray}
 \left.\min_{\{b_i;b_\alpha\}} \sum_i h_i b_i(1)\right|_{
 \left\{\begin{array}{c}
 \forall i:\ \ 0\leq b_i(\sigma_i)\leq 1\\
 \forall i\ \ \&\ \ \forall \alpha\mbox{ s.t. }  J_{\alpha i}=1:\ \
 b_i(\sigma_i)=\sum_{{\bm\sigma}_\alpha\setminus\sigma_i}b_\alpha({\bm\sigma}_\alpha)\\
 \forall\ \ i:\ \  \sum_{\sigma_i} b_i(\sigma_i)=1
 \end{array}\right.}
\label{LPL1}
\end{eqnarray}

It is straightforward to verify that the beliefs associated with
$\alpha$ could be completely removed from Eq.~(\ref{LPL1}), and the
LP problem can be restated solely in terms of the $i$-related
variables, $\beta_i\equiv b_i(1)=1-b_i(0)$.

Let us illustrate this point on example of a single $\alpha$
constraint with $m_\alpha=2$ and $q_\alpha=3$. Then the set of
allowed $\alpha$-beliefs are
\begin{equation}
 b_\alpha(1,1,0),b_\alpha(1,0,1),b_\alpha(0,1,1), \label{b23_1}
\end{equation}
and the respective set of relations (\ref{compL}) between
$\beta_1,\beta_2,\beta_3$ associated with the check $\alpha$ and the
$\alpha$ beliefs are
\begin{equation}
\hspace{-2.6cm}\beta_1=b_\alpha(1,1,0)+b_\alpha(1,0,1),\ \
\beta_2=b_\alpha(1,1,0)+b_\alpha(0,1,1),\ \ \beta_3=b_\alpha(1,0,1)+b_\alpha(0,1,1).
\label{b23_2}
\end{equation}
On the other hand the normalization condition, restated in terms of
the $\alpha$-beliefs (\ref{b23_1}), is
\begin{eqnarray}
b_\alpha(1,1,0)+b_\alpha(1,0,1)+b_\alpha(0,1,1)=1. \label{b23_3}
\end{eqnarray}
Summing Eqs.~(\ref{b23_2}) and accounting for Eq.~(\ref{b23_3}), one
finds
\begin{eqnarray}
\beta_1+\beta_2+\beta_3=2. \label{b23_4}
\end{eqnarray}
In general, one finds that the relation between $\beta$ variables
associated with an $\alpha$-constraint is
\begin{eqnarray}
\sum_i J_{\alpha i}\beta_i=m_\alpha.\label{beta_rel}
\end{eqnarray}

One derives that Eq.~(\ref{LPL1}) reduces to a simpler LP-B problem
stated solely in terms of the $\beta$ variables
\begin{eqnarray}
 \left.\min_{\beta_i} \sum_i h_i \beta_i\right|_{
 \left\{
 \begin{array}{c}
 \forall i:\ \ 0\leq \beta_i\leq 1\\
 \forall \alpha:\ \ \sum_i J_{\alpha i}\beta_i=m_\alpha\end{array}\right.}.
\label{LPL2}
\end{eqnarray}

Furthermore, one observes that, up to re-definition of $\beta_i$ to
$s_i$, Eq.~(\ref{LPL2}) is equivalent to Eq.~(\ref{LP1}). In other
words, we just showed that the $T\to 0$ solution of the BP
equations, understood as the global minimum of the Bethe free
energy, is tight,  i.e. it gives exactly the ML solution of the
binary model (\ref{ps1}).

As a side remark, one notes that it is suggestive to start exploration of the Bethe Free Energy at finite $T$ from the LP solution discussed above. It might be especially useful to initiate BP with the (easy to get) LP solution when the Bethe Free Energy optimization at finite $T$ is non-convex.

\section{Summary and Path Forward}

In this work we discussed easy problems when a zero-temperature BP scheme generates exact ML result. 
We argued that this special feature of BP is due to the fact that the related LP optimization is tight (i.e. the LP outputs ML solution as well). Our consideration was  based on the flexibility and convenience provided by the so-called Bethe Free Energy formulation, naturally relating BP and LP. The results were illustrated on two examples, FRFI model and perfect matching model. Also, we briefly discussed a broader class of easy examples related to LP with TUM matrix of constraints.

We conclude briefly mentioning some future challenges which follow from our analysis.
\begin{itemize}

\item It is useful to continue further exploration of other
models of statistical inference with loops allowing computationally
efficient optimal solutions. Thus, it would be interesting to find
examples of ``easy" non-binary problems, also these which allow
efficient and optimal finite temperature evaluation of marginal
probabilities or partition function. In this context, one mentions
exactness of BP marginals at any temperature known to hold for continuous variable Gaussian model on an arbitrary graph
\cite{01WF_Gaussian,06MJW} and also recently established, $T\to 0$, relation between an iterative algorithm
of BP type and Quadratic optimization problem \cite{06MV}.

\item Probably the most intriguing future challenge is to analyze problems
that are not computationally easy, but still close, in some metric,
to easy problems. Thus, the models discussed above, however
considered at finite, not zero, temperature may not allow an
explicit efficient solution. Similarly, perturbation of the FRFI
model with some number of graph local frustrations (e.g. some number
of randomly thrown negative $J_{ij}$ violating the TUM-feature of
the model) sets another ``close to easy" problem of theoretical and
applied interest. As suggested in \cite{05KW}, BP can be utilized as
an efficient heuristic in these ``close to easy" cases.
Note, that in this case finding minima of the Bethe Free energy may be a challenge, and the problem
turns into the quest of devising an efficient algorithms for the optimization of non-convex functions
\cite{02Yui,03YR}.
Here novel BP-convexifixation ideas developed in \cite{03WJW,05KW,07WYM,07GJ} might be helpful.
Notice also, that the loop calculus approach of \cite{06CCa,06CCb} is another useful
tool which may come handy in perturbative and non-perturbative analysis of these ``close to easy" problems.

\item BP is the algorithm of choice for decoding of
error-correction codes stated in terms of sparse graphs
\cite{63Gal}. On the other hand, the above discussion suggests that
for BP to decode optimally, or close to optimally, the graphical
structure should not necessarily be sparse. Therefore, an intriguing
question is: can one design a class of dense codes decoded
optimally (or close to optimally) by an algorithm of BP type?

\end{itemize}

The author acknowledges inspiring discussions with V. Chernyak, M. Vergassola, D.
Shah, B. Shraiman and M. Wainwright. The work was carried out under the auspices of
the National Nuclear Security Administration of the U.S. Department of Energy at Los
Alamos National Laboratory under Contract No. DE-AC52-06NA25396. The author also
acknowledges the Weston Visiting Professorship Program supporting his stay at the
Weizmann Institute, where the work was completed.

\section*{Bibliography}

\bibliographystyle{plain}
\bibliography{BP}

\begin{thebibliography}{10}

\bibitem{07BBCZ}
M.~Bayati, C.~Borgs, J.~Chayes, and R.~Zecchina.
\newblock Belief-propagation for weighted b-matchings on arbitrary graphs and
  its relation to linear programs with integer solutions, 2007.

\bibitem{08BSS}
M.~Bayati, D.~Shah, and M.~Sharma.
\newblock Max-product for maximum weight matching: Convergence, correctness,
  and lp duality.
\newblock {\em IEEE Transactions on Information Theory}, 54(3):1241--1251,
  2008.
\newblock Proc. IEEE Int. Symp. Information Theory, 2006.

\bibitem{35Bet}
H.A. Bethe.
\newblock Statistical theory of superlattices.
\newblock {\em Proceedings of Royal Society of London A}, 150:552, 1935.

\bibitem{02BH}
E.~Boros and P.~L. Hammer.
\newblock Pseudo-boolean optimization.
\newblock {\em Discrete Applied Mathematics}, 123:155--225, 2002.

\bibitem{06CCa}
M.~Chertkov and V.~Chernyak.
\newblock Loop calculus in statistical physics and information science.
\newblock {\em Physical Review E}, 73:065102(R), 2006.

\bibitem{06CCb}
M.~Chertkov and V.~Chernyak.
\newblock Loop series for discrete statistical models on graphs.
\newblock {\em Journal of Statistical Mechanics}, page P06009, 2006.

\bibitem{05FWK}
J.~Feldman, M.~Wainwright, and D.R. Karger.
\newblock Using linear programming to decode binary linear codes.
\newblock {\em Information Theory, IEEE Transactions on}, 51:954, 2005.

\bibitem{63Gal}
R.G. Gallager.
\newblock {\em Low density parity check codes}.
\newblock MIT PressCambridhe, MA, 1963.

\bibitem{07GJ}
A.~Globerson and T.~Jaakola.
\newblock Fixing max-product: Convergent message-passing algorithms for map
  lp-relaxations.
\newblock In {\em Proceedings of NIPS}, 2007.

\bibitem{02HR}
A.~Hartmann and H.~Rieger.
\newblock {\em Optimization Algorithms in Physics}.
\newblock Wiley VCH, Berlin, 2002.

\bibitem{07HJ}
B.~Huang and T.~Jebara.
\newblock Loopy belief propagation for bipartite maximum weight b-matching.
\newblock In {\em In proceedings of Artificial Intelligence and Statistics
  (AISTATS)}, 2007.

\bibitem{05KW}
V.~Kolmogorov and M.J. Wainwright.
\newblock On the optimality of tree-reweighted max-product message-passing.
\newblock In {\em Uncertainty on Artificial Intelligence}, Edinburgh, Scotland,
  2005.

\bibitem{06MJW}
D.~M. Malioutov, J.~K. Johnson, and A.~S. Willsky.
\newblock Walk-sums and belief propagation in gaussian graphical models.
\newblock {\em Journal of Machine Learning Research}, 7:2031--2064, 2006.

\bibitem{06MV}
C.~C. Moallemi and B.~Van Roy.
\newblock Convergence of the min-sum message passing algorithm for quadratic
  optimization, 2006.

\bibitem{98PS}
H.~Papadimitriou and I.~Steiglitz.
\newblock {\em Combinatorial Optimization: Algorithms and Complexity}.
\newblock Dover, 1998.

\bibitem{88Pea}
J.~Pearl.
\newblock {\em Probabilistic Reasoning in Intelligent Systems: Networks of
  Plausible Inference}.
\newblock San Francisco: Morgan Kaufmann Publishers, Inc., 1988.

\bibitem{36Pei}
H.A. Peierls.
\newblock Ising's model of ferromagnetism.
\newblock {\em Proceedings of Cambridge Philosophical Society}, 32:477--481,
  1936.

\bibitem{07SMW}
S.~Sanghavi, D.M. Malioutov, and A.~Willsky.
\newblock Linear programming analysis of loopy belief propagation for weighted
  matching.
\newblock In {\em Proceedings of NIPS}, 2007.

\bibitem{03WJ}
M.~J. Wainwright and M.~I. Jordan.
\newblock Graphical models, exponential families, and variational inference.
\newblock Technical Report 649, UC Berkeley, Department of Statistics, 2003.

\bibitem{03WJW}
M.J. Wainwright, T.S. Jaakkola, and A.S. Willsky.
\newblock Tree-based reparametrization framework for approximate estimation on
  graphs with cycles.
\newblock {\em Information Theory, IEEE Transactions on}, 49(5):1120--1146,
  2003.

\bibitem{01WF_Gaussian}
Y.~Weiss and W.T. Freeman.
\newblock Correctness of belief propagation in gaussian graphical models of
  arbitrary topology.
\newblock {\em Neural Computation}, 13(10):2173--2200, 2001.

\bibitem{07WYM}
Y.~Weiss, C.~Yanover, and T.~Melzer.
\newblock Map estimation, linear programming and belief propagation with convex
  free energies.
\newblock In {\em Proceedings of UAI}, 2007.

\bibitem{05YFW}
J.~S. Yedidia, W.~T. Freeman, and Y.~Weiss.
\newblock Constructing free-energy approximations and generalized belief
  propagation algorithms.
\newblock {\em Information Theory, IEEE Transactions on}, 51(7):2282--2312,
  2005.

\bibitem{01YFW}
J.S. Yedidia, W.T. Freeman, and Y.~Weiss.
\newblock {\em Generalized belief propagation}, volume~13, pages 689–--695.
\newblock Cambridge, MA, MIT Press, 2001.

\bibitem{02Yui}
A.~L. Yuille.
\newblock Cccp algorithms to minimize the bethe and kikuchi free energies:
  convergent alternatives to belief propagation.
\newblock {\em Neural Comput.}, 14(7):1691--1722, 2002.

\bibitem{03YR}
A.~L. Yuille and Anand Rangarajan.
\newblock The concave-convex procedure.
\newblock {\em Neural Comput.}, 15(4):915--936, 2003.

\end{thebibliography}

\end{document}